\documentclass [aps,prb,twocolumn,groupedaddress,showpacs] {revtex4}

\usepackage {graphicx,amssymb,amsbsy,color}
\begin {document}

\title {Connecting wave functions at a three-leg junction of
one-dimensional channels}
\author {Khee-Kyun Voo,$^{1, \ast}$ Shu-Chuan Chen,$^1$ Chi-Shung
Tang,$^2$ and Chon-Saar Chu$^1$}
\affiliation {$^1$Department of Electrophysics, National Chiao Tung
University, Hsinchu 30010, Taiwan, Republic of China \\ $^2$Physics
Division, National Center for Theoretical Sciences, Hsinchu 30013,
Taiwan, Republic of China}

\date {\today}

\begin {abstract}

We propose a scheme to connect the wave functions on different
one-dimensional branches of a three-leg junction (Y-junction). Our scheme
differs from that due to Griffith [Trans. Faraday Soc. {\bf 49}, 345
(1953)] in the respect that ours can model the difference in the widths of
the quasi-one-dimensional channels in different systems. We test our
scheme by comparing results from a doubly-connected one-dimensional system
and a related quasi-one-dimensional system, and we find a good agreement.
Therefore our scheme may be useful in the construction of one-dimensional
effective theories out of (multiply-connected) quasi-one-dimensional
systems.

\end {abstract}

\maketitle

\section {Introduction}

For a system which comprises quasi-one-dimensional (Q1D) channels, when
only the low-energy regime at near the first subband bottom is
considered, it can usually be modeled by a one-dimensional (1D) system.
When the system is multiply-connected and consists of multi-leg junctions,
the wave functions on the branches are usually connected at the junctions
by the Griffith scheme, \cite {Kuh49,Gri53,RS53,KS99} the Shapiro scheme,
\cite {Sha83,BIA84,ES89} or schemes
alike. Since such formulations greatly reduce the calculational effort of
complicated multiply-connected mesoscopic systems, they have been used
widely in the literature. For example, see Refs.~ \onlinecite
{GIA84,Xia92,MHZ93,DJ94,Mos97,RC98,BJ03,JSJ01,MPV04,BGC04,FMB05,AWS05}
and the references therein. However, arguments which lead to these 
connecting schemes are kinematical, \cite
{Kuh49,Gri53,RS53,KS99,Sha83,BIA84,ES89} and it is
not clear what kind of junction in practice they describe. Moreover, a
comparison between the results of these schemes and that of the exact
calculation of Q1D systems has never been done. It is the purpose of this
paper to make a comparison between the Griffith result, the Q1D result,
and the result due to a scheme we propose in this paper. We find that for
clean junctions of Q1D channels, the Griffith result is not even
qualitatively in accord with the exact result. The scheme we derive gives
a result that compares much better with the exact result.

At a $N$-leg junction of 1D channels, the wave function continuity
condition is a requirement that must be respected. Besides, the Griffith
scheme \cite {Kuh49,Gri53,RS53,KS99} demands that the sum of the
derivatives of the wave functions on the different branches at the
junction is zero, i.e.,
\begin {eqnarray}
\displaystyle \sum_{i = 1} ^N { {\partial \psi_i} \over {\partial x_i} }
= 0,
\label {gri}
\end {eqnarray}
where the directions of the coordinates are defined either diverging
from or converging to the junction. This is the simplest way to impose
the unitarity condition of no net current flows into the junction, i.e.,
$\sum_i {\rm Re} ~ \psi_i^\ast (-i \partial \psi_i / \partial x_i ) = 0$.
When there is a magnetic field, the requirement is rephrased as the sum of
the covariant derivatives is zero, i.e., $\sum_{i} ( \partial /
\partial x_i - i e A_i ^\parallel ) \psi_i = 0$, where $A_i ^\parallel$ is
the component of the vector potential parallel to branch $i$ at the
junction. On the other hand, the Shapiro scheme \cite {Sha83,BIA84,ES89}
directly demands that the scattering matrix connecting the in-going and
out-going waves at the junction be unitary, and a general matrix with
free parameters is written down. When the spin degree of freedom is
considered, these schemes are straightforwardly applied to each spin
channel. \cite {JSJ01,MPV04,BGC04,FMB05,AWS05} These schemes and the likes
have been taken for granted and used widely in the literature.

\section {Formulations and models}

\begin {figure}


\includegraphics{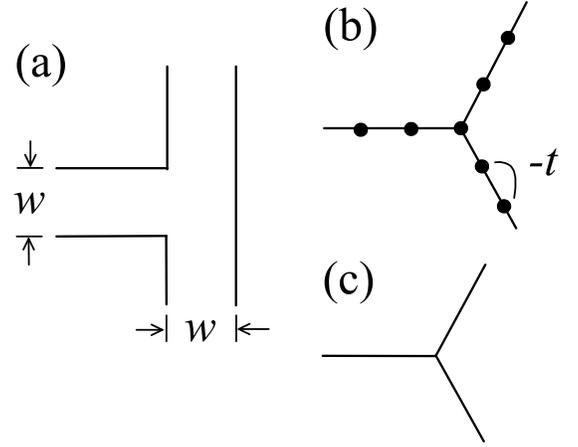}
\vspace {6cm}

\caption {
(a) The original Y-junction of Q1D channels considered in this paper, (b)
the reformulated Y-junction of tight-binding channels, and (c) the
effective Y-junction of 1D channels affiliated with the Griffith or our
connection scheme.
}
\label {junct}
\end {figure}

We approach the problem from another point of view. For a Q1D system with
equal-width channels (the ``width'' is an ill-defined quantity in snaking
channels but nevertheless we may talk about it when the curvatures are
small enough), we may approximate a three-leg junction (Y-junction) and
its branches [e.g., see Fig.~\ref {junct}(a)] by a tight-binding (TB) 
model as shown in Fig.~\ref {junct}(b). \cite {VCTC05a}

The tight-binding model is described by a second quantised Hamiltonian
\begin {eqnarray}
H = \sum _{\bf i,j} c_{\bf i}^\dagger h_{\bf ij} c_{\bf j}, 
\end {eqnarray}
where $c_{\bf i}$ is the annihilation operator of a spinless particle on
site ${\bf i}$, and $h_{\bf ij}$ is a matrix element which is complex in
general. The element $h_{\bf ij}$ is called a hopping when ${\bf i} \neq
{\bf j}$, and an onsite potential when ${\bf i} = {\bf j}$. 
Defining a basis set $\{ |{\bf i} \rangle \}$ by $|{\bf i} \rangle \equiv
c_{\bf i}^\dagger | 0 \rangle$, where $| 0 \rangle$ is the no-particle
state, one can write the time-independent Schr$\ddot {\rm o}$dinger
equation $H | \psi \rangle = E | \psi \rangle$, where $E$ is the energy
of the particle, into the form 
\begin {eqnarray}
\sum _{\bf j} ( h_{\bf ij} - E \delta _{\bf ij} ) {\bar \psi} _{\bf j} =
0, 
\label {tb}
\end {eqnarray}
where $\bar \psi _{\bf j} \equiv
\langle {\bf j} | \psi \rangle$ is the TB wave function at site ${\bf j}$.
We define that the hopping exists only between nearest-neighbor sites, and
be denoted by $- t$. The onsite potential at site ${\bf i}$ is denoted by
$V_{\bf i} + 2t$.

The magnitude of the hopping $- t$ is obtained by the following argument.
Let a Q1D channel be approximated by a finite-difference square grid, with
three grid-points across the channel, one at the center and each edge.
Then the distance between the grid-points will be $w/2$, where $w$ is the
width of the channel, and the hopping in the finite-difference 
time-independent Schr$\ddot {\rm o}$dinger equation \cite {Dat95} will be
$- t = - 2\hbar^2/(mw^2)$. We will assume the same hopping in our TB
formulation.

Away from the junction, the TB time-independent Schr$\ddot {\rm o}$dinger
equation reads \cite {Dat95}
\begin {eqnarray}
- t (\bar \psi_{\bf i+1} - \bar \psi_{\bf i}) + t (\bar \psi_{\bf i} -
\bar \psi_{\bf i-1}) + (V_{\bf i} - E) \bar \psi_{\bf i} = 0, 
\end {eqnarray}
where $E$ is the energy. In the long-wavelength limit it reduces, as it
should, to the 1D second order differential time-independent Schr$\ddot
{\rm o}$dinger equation, $-[\hbar^2/(2m)] \partial _x ^2 \psi (x) +
[V(x)-E] \psi (x) = 0$. 

At a Y-junction, the TB time-independent Schr$\ddot {\rm o}$dinger
equation reads,
\begin {eqnarray}
\displaystyle (\bar \psi_1 - \bar \psi_0) + (\bar \psi_2 -
\bar \psi_0) + (\bar \psi_3 - \bar \psi_0) + 
{ {E-V_0+t}  \over {t} } ~ \bar \psi_0 = 0, 
\label {tby}
\end {eqnarray}
where the subscript ``0'' denotes the site at the junction, and ``1'',
``2'', and ``3'' denote the sites on the branches nearest to the site at
the junction [i.e., in Eq.~(\ref {tb}), take ${\bf i} = 0$, and ${\bf
j} = 0, 1, 2$, and 3]. It is seen that the Griffith scheme formulated in
Eq.~(\ref {gri}) is recovered only when $E-V_0+t=0$ at the junction. It is
reasonable to set $E=0$ here since we are considering energies at near the
band-bottom and $E \ll t$. But one still requires $V_0=t$ to send the last
term in Eq.~(\ref {tby}) to zero. In other words, the Griffith
connection scheme \cite {Kuh49,Gri53,RS53,KS99} actually describes a
Y-junction of Q1D channels with a {\it repulsive} potential with a
strength of the order
of $t$. Whereas in this paper we propose a connection scheme in the
long-wavelength limit for a {\it clean} Y-junction [i.e., $V_0=0$ in
Eq.~(\ref {tby})] of Q1D channels. At a Y-junction of 1D channels [see
Fig.~\ref {junct}(c)], we propose
\begin {eqnarray}
\displaystyle \sum_{i = 1} ^3 { {\partial \psi_i} \over {\partial x_i}
} + {{2 \nu} \over w } ~ \psi _1 = 0, 
\label {voo}
\end {eqnarray}
where the directions of the coordinates are defined to be diverging from
the junction. If Eq.~(\ref {voo}) is reached by dividing Eq.~(\ref {tby})
by $w/2$ and letting $w \rightarrow 0$ (remember that $E/t$ and $V_0/t$
have been set to zero), the factor $\nu$ will be equal to 1. 
Adopting $\nu = 1$ indeed results in a good enough qualitative comparison
with the Q1D result. But we will see that choosing $\nu \simeq 1.9$ may
bring the 1D and Q1D results to a semi-quantitative agreement, which means
that the term has been underestimated. The TB argument serves to bring out
the $1/w$ dependence of the term, and the fixing of $\nu$ will be
discussed using concrete examples. The effect of the channel width  is
hence included, in contrast to the Griffith scheme (the case of $\nu=0$).
The $1/w$ dependence results in an effect that is more prominent at
smaller channel widths, and this understanding may also help to relate
studies on the quantum graph theory \cite {KS99,ES89} to the
practical experiments. The case of a general $N$-leg junction can also be
worked out likewise.

\begin {figure}


\includegraphics{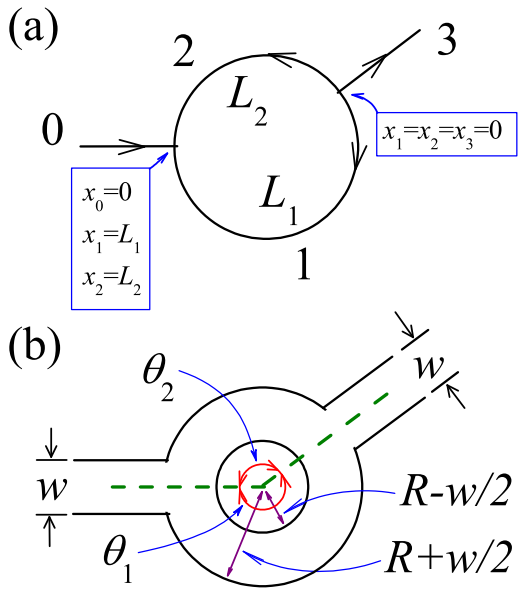}
\vspace {8cm}

\caption {
(Color online) The doubly-connected 1D system and the related Q1D system
we consider. (a) The 1D system is a ring (with arms labeled by 1 and 2)
connected to two leads (labeled by 0 and 3). The coodinate system $x_i$ is
defined for the line segment labeled by $i$ ($i=$ 0, 1, 2, and 3). The
arrows denote the positive directions of the coordinates, the right
Y-junction is defined at $x_1=x_2=x_3=0$, and the left Y-junction at
$x_0=0$, $x_1=L_1$, and $x_2=L_2$. (b) The Q1D system is an annulus with
two radially connected leads. Both the annulus and the leads have the same
width.
}
\label {ring}
\end {figure}

In this paper we compare the Griffith and our schemes with the exact Q1D
calculation in a chosen type of system. We calculate the transmission
probability for a 1D ring connected to two leads [see Fig.~ \ref
{ring}(a)], which is the simplest multiply-connected 1D system, using the
Griffith and our connection schemes at the Y-junctions. In addition, we
also calculate the transmission probability for a similar system, an
annulus connected to two Q1D leads [see Fig.~\ref {ring}(b)], using the
exact mode-matching method. The two three-leg junctions in Fig.~\ref
{ring}(b) resemble the one in Fig.~\ref {junct}(a). Note that the
transmission probability is directly related to the experimentally
measurable conductance. \cite {Dat95} We will sketch how we have done the
calculations, and we refer the readers to the literatures for more 
details.

In a 1D model as shown in Fig.~ \ref {ring}(a), the wave function on each
line segment at a given positive energy $E$ is a superposition of forward
and backward traveling waves, i.e., 
\begin {eqnarray}
\psi_i (x_i) = A_i e^{ik x_i} + B_i e^{-ik x_i}, ~~~~ i = 0, 1, 2, {\rm 
and} ~ 3, 
\label {psi}
\end {eqnarray}
where $k \equiv {\sqrt {2mE} } / \hbar$ and $m$ is the effective mass of
the particle. The wavelength $\lambda$ is given by $\lambda \equiv 2 \pi /
k$. The $x_{0,1,2,3}$ are the coordinates on the line segments 
correspondingly, and the coordinates have positive directions as that 
defined in Fig.~\ref {ring}. 
We define $x_1 = x_2 = x_3 = 0$ at the right junction, and $x_0 = 0$, $x_1
= L_1$, $x_2 = L_2$ at the left junction, where $L_{1,2}$ are the lengths
of the arms between the junctions [see Fig.~\ref {ring}(a)]. 
The $A_i$ ($B_i$) is the coefficient of a forward (backward) traveling
wave. Since we consider particles incident from the left, we set $A_0=1$
and $B_3=0$. Then the continuity requirement 
\begin {eqnarray}
\psi_0 | _{x_0=0} = \psi_1 | _{x_1=L_1} = \psi_2 | _{x_2=L_2} 
\end {eqnarray}
and
\begin {eqnarray}
\psi_1 | _{x_1=0} = \psi_2 | _{x_2=0} = \psi_3 | _{x_3=0}, 
\end {eqnarray}
and the Griffith unitarity imposition [following Eq.~(\ref {gri})]
\begin {eqnarray}
\left. { {\partial \psi_0} \over {\partial x_0} } \right| _{x_0 = 0} +
\left. { {\partial \psi_1} \over {\partial x_1} } \right| _{x_1 = L_1} +
\left. { {\partial \psi_2} \over {\partial x_2} } \right| _{x_2 = L_2} = 0
\end {eqnarray}
and 
\begin {eqnarray}
\left. { {\partial \psi_1} \over {\partial x_1} } \right| _{x_1 = 0} + 
\left. { {\partial \psi_2} \over {\partial x_2} } \right| _{x_2 = 0} +
\left. { {\partial \psi_3} \over {\partial x_3} } \right| _{x_3 = 0} = 0
\end {eqnarray}
constitute an equation set which contains six equations with the six
unknowns $\{B_0;A_1,B_1;A_2,B_2;A_3\}$ which have been defined in
Eq.~(\ref {psi}). Hence
the transmission amplitude $A_3$ can be solved, and the transmission
probability $T = |A_3|^2$ be found. We may also replace the Griffith
unitarity condition by our unitarity condition [following Eq.~(\ref
{voo})] 
\begin {eqnarray}
\left. { {\partial \psi_0} \over {\partial x_0} } \right| _{x_0 = 0} +
\left. { {\partial \psi_1} \over {\partial x_1} } \right| _{x_1 = L_1} +
\left. { {\partial \psi_2} \over {\partial x_2} } \right| _{x_2 = L_2} +
\left. {{2 \nu} \over w} ~ \psi_0 \right| _{x_0 = 0} = 0 
\end {eqnarray}
and
\begin {eqnarray}
\left. { {\partial \psi_1} \over {\partial x_1} } \right| _{x_1 = 0} +
\left. { {\partial \psi_2} \over {\partial x_2} } \right| _{x_2 = 0} +
\left. { {\partial \psi_3} \over {\partial x_3} } \right| _{x_3 = 0} +
\left. {{2 \nu} \over w} ~ \psi_3 \right| _{x_3 = 0} = 0, 
\end {eqnarray}
and too the transmission probability can be solved. We will discuss the
fixing of $\nu$ later in this paper.

Besides the mentioned 1D model, we also solve a related Q1D model in a way
as that of Xia and Li in Ref.~\onlinecite {XL02}. Consider an annulus with
an inner and an outer radii of $R-w/2$ and $R+w/2$ respectively, and two
leads of width $w$ radially connected to it as shown in Fig.~\ref
{ring}(b). The wave function is governed by the two-dimensional (2D)
differential time-independent Schr$\ddot {\rm o}$dinger equation. 
In a lead it can be expanded in terms of transverse modes (subbands)
and longitudinal forward and backward modes, i.e., $\psi_{\rm lead} (x,y) 
= \sum _{l=1} ^{N} (a_l e^{ik_lx} + b_l e^{-ik_lx}) {\rm sin} (l \pi y /
w)$, where $x$ and $y$ are respectively the longitudinal and transverse
coordinates for the lead. The $k_l$ and $l$ are related by $k_l^2 +
(l\pi/w)^2 = 2mE/\hbar^2$, where $E$ is the energy (positive) of the
particle, and $k_l$ can be real or imaginary.
In the annulus the wave function can be expanded by radial and angular
modes, i.e., $\psi_{\rm annulus} (r,\theta) = \sum _{l=-M} ^M \phi_l (kr)
e^{il\theta}$, where a radial mode is given by $\phi_l (kr) \equiv c_l J_l
(kr) + d_l Y_l (kr)$, and $k = \sqrt {2mE}/\hbar$. The $r$ and $\theta$
are the radial and angular coordinates respectively; and the $J_l$ and
$Y_l$ are the Bessel functions of the first and second kinds
respectively.
We demand $\phi_l | _{r=R-w/2} = 0$ for any $\theta$, $\psi_{\rm
annulus} | _{r=R+w/2} = 0$ when $\theta$ is away from the leads, but 
$\psi_{\rm annulus} | _{r=R+w/2} = \psi_{\rm lead}$ when $\theta$ is in
the range of a lead. Also, the radial derivative $\partial \psi_{\rm
annulus} / \partial r$ is equated with the longitudinal derivative
$\partial \psi_{\rm lead} / \partial x$ when they meet at the outer arc of
the annulus. The difference between the straight transverse cuts of the
leads and the outer arcs of the annulus is neglected. The wave functions
in the leads and the annulus are hence matched.
Expanding the wave functions in different regions with sufficient numbers
of modes, \cite {VCTC05e} one gets a set of equations relating the
coefficients of the modes in different regions. With a given energy $E$
and specified in-going subbands, one can obtain the transmission
probabilities in the out-going subbands. 

In the Q1D case, we will consider that the particle is incident from one
lead, and its energies is below the second subband and hence the particle
propagates only within the first subband. The resulting transmission
probability is to be compared with that in the 1D case.
We will use the more convenient longitudinal wave number $k _\parallel
\equiv \sqrt {2m (E - E_{0} ^{\rm ann.})} / \hbar$ instead of the energy
$E$, where $E_{0} ^{\rm ann.}$ is the energy of the nodeless ground state
of the isolated annulus in an individual case. \cite {VCTC05b}  
Defining a longitudinal wavelength $\lambda _ \parallel$ by $\lambda _
\parallel \equiv 2 \pi / k _\parallel$ implies that the long-wavelength
limit we consider is at $\lambda _\parallel \gg w$. Here we define the
arm lengths by $L_{1,2} \equiv R  \theta_{1,2}$, where $\theta_{1,2}$ are
the angles shown in Fig.~\ref {ring}(b). 

\section {Comparison between results}

\begin {figure} 


\includegraphics{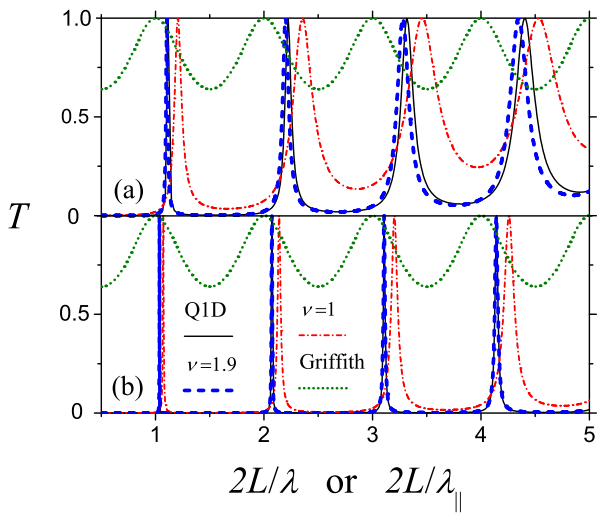}
\vspace {8cm}

\caption {
(Color online) The transmission probability $T$ is plotted versus the
dimensionless longitudinal wave numbers ($2L / \lambda _\parallel$ in the
case of Q1D channels, and $2L / \lambda$ in the case of 1D channels), for
the case of $L_1 = L_2 \equiv L$.
The Q1D results (solid lines), 1D results due to our scheme [dashed
($\nu=1.9$) and dash-dotted ($\nu=1$) lines], and 1D results due to the
Griffith scheme (dotted lines) are shown. $T$ is plotted for the cases
of broad and narrow Q1D channels, (a) $R/w=3.5$ and (b) $R/w=9.5$.
Note that the Griffith result is independent of the channel widths.
In the narrow channel case [(b)], the difference between the Q1D and
$\nu = 1.9$ results is indiscernible in the scale of this graph.
}
\label {nofano}
\end {figure}

Figure \ref {nofano} shows the transmission probabilities obtained by  
different schemes, for the case of symmetrical arms in the ring ($L_1 =
L_2 \equiv L$). We have considered broad [Fig.~\ref {nofano}(a)] and
narrow [Fig.~\ref {nofano}(b)] channels, and in both cases we have
presented the result of the Q1D calculation, the result of the Griffith
scheme, and the results of our scheme (with $\nu=1$ and 1.9).
The Griffith result is seen to differ very much from the Q1D result in all
cases. The $\nu=1$ scheme qualitatively captures the trend of change in
the Q1D result when the channel width is changed, while the $\nu=1.9$
scheme captures the Q1D result most satisfactorily. 

Besides the Griffith result, it is seen that all results in Fig.~\ref
{nofano} show Breit-Wigner (BW) resonance peaks. \cite {BW36} These BW
peaks become sharper and shift toward the left
when the channels narrow down, i.e., $w/R \rightarrow 0$ [compare 
Figs.~\ref {nofano}(a) and (b)]. Those peaks are due to the quasibound
levels in the arms, and they are seen to be always blue shifted \cite
{VCTC05d} from the exact levels. In the 1D case, the exact levels are at
$2L / \lambda \equiv k L / \pi = {\rm integer}$. 
The quasibound levels and the blue shift are results of the presence of an
{\it attractive} potential at a Y-junction. \cite {VCTC05c} While the
attractive potential in our scheme is manifest [see Eq.~(\ref {voo})], the
potential at a Y-junction of Q1D channels is not so obvious, but can have
an intuitive understanding as follows. Since a particle feels less
confined at near a
junction, the ``band-bottom'' at the vicinity of a junction is effectively
lower, and therefore the region acts as an attraction center.
This potential becomes stronger when the channels become narrower, and
that leads to the sharper and less blue shifted BW peaks [compare
Figs.~\ref {nofano}(a) and (b)]. 
The growth of the potential at narrowing channels can be understood as
a result of the departure from the case of very broad channels (i.e., $w
\sim R$), in which the system has no difference in the ``band-bottom''
everywhere.

While the result from the Griffith scheme is independent of the channel
width and disagrees with the Q1D result, our scheme captures the trend of
change in the transmission probability when the channel width is varied. 
Therefore, our scheme has correctly included the attractive nature of the
clean Y-junction of Q1D channels, though the strength has been 
underestimated (i.e., $\nu = 1.9$ is prefered to $\nu = 1$). 
The misjudgment of an appropriate value for the parameter $\nu$ is due to
the fact that the details of the shape of the Y-junctions of Q1D channels 
and the actual dimensionality of the channels are relevant.
For instance, our simple TB argument which leads to Eqs.~(\ref {tby})
and (\ref {voo}) does not show the difference between junctions with
different relative directions of branching channels, and also does not
distinguish a three-dimensional (3D) cylinder from a 2D strip as a Q1D
channel. But in reality, the appropriate parameter $\nu$'s in those
different cases may likely be different. 
In the 2D cases we have just seen in Fig.~\ref {nofano}, the same kind of
Y-junction has been involved, and the effective potential in our scheme is
characterized by an almost constant $\nu$ in both the broad [Fig.~\ref
{nofano}(a)] and narrow [Fig.~\ref {nofano}(b)] channel cases.

\begin {figure}


\includegraphics{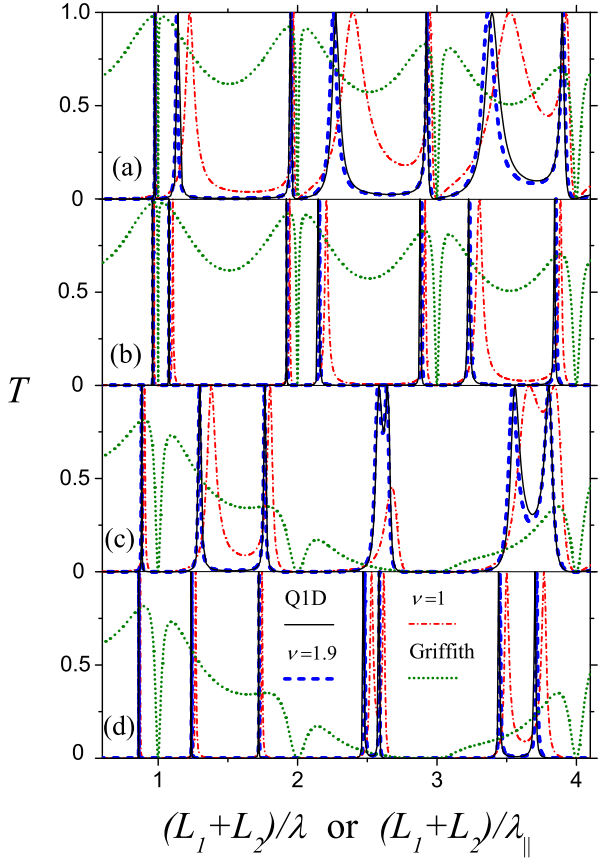}
\vspace {13cm}

\caption {
(Color online) The transmission probability $T$ is plotted versus the
dimensionless longitudinal wave numbers [$(L_1+L_2) / \lambda _\parallel$
in the case of Q1D channels, and $(L_1+L_2) / \lambda$ in the case of 1D
channels], for the case of asymmetrical arm lengths. 
The Q1D results (solid lines), 1D results due to our scheme [dashed
($\nu=1.9$) and dash-dotted ($\nu=1$) lines], and 1D results due to the
Griffith scheme (dotted lines) are shown.
$T$ is plotted for the cases of broad and narrow Q1D channels, with small
and appreciable differences in the arm lengths. $T$ is plotted for (a)
$R/w=3.5$, $L_2/L_1=0.9$, (b) $R/w=9.5$, $L_2/L_1=0.9$, (c) $R/w=3.5$,
$L_2/L_1=0.7$, and (d) $R/w=9.5$, $L_2/L_1=0.7$.
Note that the Griffith result is independent of the channel widths.
In the narrow channel cases [(b) and (d)], the differences between the Q1D
and $\nu = 1.9$ results are indiscernible in the scale of this graph.
}
\label {withfano}
\end {figure}

Therefore, though the parameter $\nu$ can not be derived analytically, it
can be readily fixed for a particular kind of junction by comparing the 1D
result with the Q1D result. What we have done in Fig.~\ref {nofano} has
been a comparison which involves a tedious calculation. Actually, other
simpler comparisons also work. For instance, one may consider the bound
state at the junction due to the attraction. \cite {SRW89} 
On one hand, for a junction of three 1D channels, with the channels
extended to infinity like what we depict in Fig.~\ref {junct}(c), the
negatively valued bound state energy $E$ can be readily found by using
$\psi_i = e ^{-\kappa x_i}$, where $i =$ 1, 2, 3, $\kappa \equiv \sqrt
{-2mE} / \hbar$, and Eq.~(\ref {voo}). The energy $E$ is found to be lower
than zero by an amount of $2 \nu^2 \hbar^2 / (9 m w^2)$. 
On the other hand, the bound state at a T-shaped junction of three Q1D
channels, with the channels extended to infinity like what we depict in
Fig.~\ref {junct}(a), was studied by Schult $et~al.$ \cite {SRW89} The
energy of the state was numerically found to be lower than the first
subband bottom by an amount of $0.19 \pi^2 \hbar^2 / (2mw^2)$. \cite
{SRW89} Equating the two energies in the 1D and Q1D cases, one gets $\nu
\simeq 2.05$, which is about the number we use in Fig.~\ref {nofano}, and
as we will see, that in Fig.~\ref {withfano}.

Figure \ref {withfano} shows the transmission probabilities for the case
of asymmetrical arm lengths. It is seen that in all cases, broad channels
[Figs.~\ref {withfano}(a) and (c)] and narrow channels [Figs.~\ref
{withfano}(b) and (d)], small difference in arm lengths [Figs.~\ref
{withfano}(a) and (b)] and appreciable difference in arm lengths
[Figs.~\ref {withfano}(c) and (d)], there are good comparisons between the
results due to our $\nu=1.9$ scheme and the Q1D calculation. All the
essential features, such as the relative positions of the BW and Fano
profiles \cite {Fan61} in the Q1D results, are nicely reproduced. Note
that the number 1.9 agrees with the one used in Fig.~\ref {nofano}.

For the 1D models, including Griffith's and ours, the perfectly zero
transmission dips of the Fano profiles are located exactly at the
eigenenergies of an isolated ring \cite {VC05} with a circumference of
$L_1+L_2$. In the 1D case, these eigenenergies are exactly at $(L_1+L_2) /
\lambda \equiv k (L_1+L_2) / (2 \pi) = {\rm integer}$. In the Q1D cases
shown in Figs.~\ref {nofano} and \ref {withfano}, the eigenenergies are
numerically found to be at the $(L_1+L_2) / \lambda _\parallel$'s deviated
by not more than 0.5$\%$ from the integers on the horizontal axes. 
For the Q1D model, we find that those zero transmission dips may coincide
with the eigenenergies of an isolated annulus only in the long-wavelength
limit. 
As in the case of symmetrical arms, the Griffith result disagrees with
the Q1D result, and our simple TB argument which leads to Eqs.~(\ref
{tby}) and (\ref {voo}) has underestimated the strength of the effective
potential at the junction, i.e., $\nu=1.9$ is prefered to $\nu=1$. 

\section {Concluding remarks}

%
It is seen that in all the above cases the results from the Griffith
scheme are not in congruence with the Q1D results. The Griffith result
is regardless of the channel width, whereas the Q1D result shows a
strong dependence on that. Our model gives a result in much better
agreement with the Q1D result. The trend of change in the transmission
probability and the relative positions of the resonance profiles are
impressively reproduced. 
In view of these calculations, it is clear that the Griffith scheme which
is frequently adopted in the literature, does not describe a clean 
junction of Q1D channels and is definitely not for the 1D limit of the Q1D
models. In the small width limit, a Y-junction of Q1D channels is a
strong scatterer, and that makes
the Q1D system studied in this paper not at all an ``open'' system.
Speaking reversely, adding a repulsive potential to a Y-junction of Q1D
channels may weaken the scattering effect and enhance the transmission
through the junction at low energies, and away from the levels.
When a strong magnetic field is present, our model may not apply since
the field creates an additional asymmetric transverse confinement.

In conclusion, we have proposed a connection scheme with a parameter
$\nu$ at a Y-junction of 1D channels. The parameter $\nu$ can be most
easily fixed by comparing the energy of the bound state at a Y-junction
of Q1D channels, to the energy of the bound state at a Y-junction of 1D
channels due to Eq.~(\ref {voo}).
The scheme reflects the presence of an effectively attractive potential at
a clean Y-junction of Q1D channels. The disregard of this potential in the
Griffith scheme makes its result compares poorly with the exact Q1D
result.

\indent {\bf Acknowledgments -} This work is supported by the National
Science Council of Taiwan under Grant No. 94-2112-M-009-017. We thank the
National Center for Theoretical Sciences of Taiwan for letting us to use
their facilities.

\end{document}